\documentclass[aps,jcp,preprint,superscriptaddress,longbibliography]{revtex4-2}

\usepackage[utf8]{inputenc}
\usepackage[T1]{fontenc}
\usepackage{amsmath,amssymb}
\usepackage{graphicx}
\usepackage{booktabs}
\usepackage{xcolor}
\usepackage{siunitx}
\usepackage{microtype}
\usepackage{hyperref}
\usepackage[capitalise]{cleveref}

\newcommand{\Pval}{P_{\mathrm{val}}}
\newcommand{\Pcap}{P_{\mathrm{cap}}}
\newcommand{\Vcup}{V_{\mathrm{cup}}}
\newcommand{\Vdimple}{V_{\mathrm{dimple}}}
\newcommand{\Vvoid}{V_{\mathrm{void}}}
\newcommand{\Vcappar}{V_{\mathrm{cap}}}
\newcommand{\vcup}{v_{\mathrm{cup}}}
\newcommand{\vdimple}{v_{\mathrm{dimple}}}
\newcommand{\vvoid}{v_{\mathrm{void}}}
\newcommand{\vcappar}{v_{\mathrm{cap}}}
\newcommand{\tdry}{\tau_{\mathrm{dry}}}
\newcommand{\twick}{\tau_{\mathrm{wick}}}
\newcommand{\dPcap}{\Delta P_{\mathrm{cap}}}
\newcommand{\dPdrain}{\Delta P_{\mathrm{drain}}}
\newcommand{\Qwick}{Q_{\mathrm{wick}}}
\newcommand{\Rduct}{\mathcal{R}}
\newcommand{\Perim}{\mathcal{P}}
\newcommand{\Cd}{\mathcal{C}}
\newcommand{\nm}{\,\mathrm{nm}}
\newcommand{\tnd}{\tilde{t}}
\newcommand{\hmax}{h_{\max}}
\newcommand{\hmin}{h_{\min}}
\newcommand{\zfl}{z_{\mathrm{floor}}}
\newcommand{\zri}{z_{\mathrm{rim}}}
\newcommand{\tr}{\varepsilon}            
\newcommand{\trc}{\varepsilon_{1}}       
\newcommand{\Cnum}{\Lambda}              

\newcommand{\numMu}{1.2\times10^{-3}}
\newcommand{\numTone}{38.7}
\newcommand{\numToneExp}{35.6}
\newcommand{\numDtone}{144}
\newcommand{\numRatioLimit}{8.4}
\newcommand{\numRatioTable}{9.0}
\newcommand{\numVcapRed}{0.032}
\newcommand{\numLambda}{7.3\times10^{6}}
\newcommand{\numTwick}{11.9}
\newcommand{\numRatioChain}{8.4\times10^{3}}
\newcommand{\numRatioMax}{5\times10^{5}}

\newcommand{\numCduct}{1.045}
\newcommand{\numCzero}{1.046}

\newcommand{\numTwickSlit}{3.2}
\newcommand{\numTwickPaper}{5.2}

\newcommand{\numPerimThirty}{95}
\newcommand{\numPoissonErr}{1.5\times10^{-5}}
\newcommand{\numDeltaCrit}{0.20}
\newcommand{\numCeff}{1.12}
\newcommand{\numStep}{2.4}
\newcommand{\numPvalLo}{0.87--0.94}
\newcommand{\numPvalHi}{0.25--0.29}
\newcommand{\numPcapTen}{0.06}
\newcommand{\numPcapThirty}{0.14}
\newcommand{\numPcapFifty}{0.72}
\newcommand{\numPcapSeventy}{0.76}

\newcommand{\numPixelDev}{1.3\times10^{-5}}
\newcommand{\numSpreadPdeMin}{16.1}
\newcommand{\numSpreadPdeMax}{24.9}
\newcommand{\numReliefPdeMin}{27}
\newcommand{\numReliefPdeMax}{68}
\newcommand{\numSlopePdeMax}{0.39}
\newcommand{\numProxyMin}{0.67}
\newcommand{\numProxyMax}{0.78}
\newcommand{\numKappa}{0.26}
\newcommand{\numDtPde}{0.2}
\newcommand{\numDtDiff}{4\times10^{-4}}
\newcommand{\numSpreadEqMin}{3.6}
\newcommand{\numSpreadEqMax}{6.7}

\newcommand{\numEqRatioMin}{0.10}
\newcommand{\numEqRatioMax}{0.13}

\begin{document}

\title{Capillary partitioning of solutes on conformally coated colloidal
	monolayers: a universal geometric threshold}

\author{Renaud A.\,L.\ Vall\'ee}
\email{renaud.vallee@crpp.cnrs.fr}
\affiliation{CRPP, CNRS, Universit\'e de Bordeaux, 115 avenue Schweitzer,
	33600 Pessac, France}
\author{Carlos Drummond}
\affiliation{CRPP, CNRS, Universit\'e de Bordeaux, 115 avenue Schweitzer,
	33600 Pessac, France}

\date{\today}

\begin{abstract}
	Depositing a conformal solid film on a hexagonal monolayer of spheres
	progressively closes the interstitial pores, and this closure controls
	where a dilute solute spin-coated from a volatile solvent ends up after
	drying: buried in the interstitial valleys or spread over the exposed
	caps. We derive the fraction of solute that ends in the valleys from first
	principles, for any sphere radius, film thickness and wetting liquid. The
	control parameter is the film thickness divided by the sphere radius, and
	a purely geometric threshold, about 0.155, separates two regimes for all
	sphere sizes. Below it, the three-sphere interstice is an open cup that
	the liquid fills through a narrow throat at the sphere equator, and about
	nine tenths of the solute end in the valleys. Above it, the throat pinches
	shut, sealing the cup and leaving only a shallow surface dimple, so this
	fraction drops more than twofold, never to recover. Treating the throat
	as a nearly triangular duct gives a closed-form filling rate, confirmed by
	exact flow calculations. During spin-coating, filling outpaces drying by
	up to five orders of magnitude, except within a fraction of a nanometre of
	the threshold. Partitioning is therefore set by capacity alone, through a
	one-parameter ratio of exact void volumes, identical for all sphere sizes. Thin-film simulations confirm the direction of liquid
	transport. For dye in ethanol on tantalum-pentoxide-coated polystyrene
	lattices, the predicted on-cap fraction rises from 0.06 to 0.76 across the
	threshold, agreeing at both ends with values inferred from measured
	Purcell factors.
\end{abstract}

\maketitle

\section{Introduction}
\label{sec:intro}

Monolayers of colloidal spheres arranged on a hexagonal lattice are a
workhorse template in surface nanofabrication. They underpin nanosphere
lithography~\cite{Hulteen1995,Haynes2001,Deckman1982} and the fabrication of
plasmonic~\cite{Fredriksson2007}, photonic~\cite{Jonsson2007}, and
fluorescence-enhancing~\cite{Falamas2026} surfaces. A common processing step is
to cover the assembled monolayer with a thin solid film---by atomic-layer
deposition (ALD)~\cite{George2010}, sputtering, or evaporation---to tune the
optical, chemical, or mechanical response of the surface. As the film thickens,
the geometry of the voids between the spheres changes qualitatively: at small
thickness the voids are open channels that connect the buried interstitial
space to the outer surface, whereas beyond a critical thickness the film
bridges the three-sphere junctions and seals the voids into isolated pockets.

A second common step is to deposit a dilute solute---a fluorescent dye, a
functional molecule, a nanoparticle---from a volatile solvent, by
spin-coating. As the solvent evaporates, surface tension pulls the residual
liquid toward regions of negative curvature, i.e.\ into the topographic
valleys~\cite{deGennes2004,Seemann2005,Bico2002}, carrying the solute with it.
Where the solute finally sits---concentrated in the valleys or smeared over the
caps---is not a cosmetic detail: for optical applications it sets the
fluorescence yield and the near-field coupling that govern single-molecule
detection~\cite{Kinkhabwala2009}, surface-enhanced Raman
scattering~\cite{Alvarez2010}, and the ensemble Purcell factor of dye emitters
on coated microsphere lattices~\cite{Falamas2026}. Yet a predictive,
first-principles description of this partitioning as a function of the coating
thickness has been missing. Recent fluorescence measurements on
PS/\mbox{Ta$_2$O$_5$} lattices show that an empirical average over cap versus
valley placements accounts for the measured Purcell factors across shell
thicknesses~\cite{Falamas2026}; the present work supplies the geometric theory
of that placement.

The physical question is simple to state. Two clocks run during drying: the
time the liquid needs to redistribute over the textured surface, and the time
the solvent needs to evaporate. If redistribution is fast, the solute
distribution is set by \emph{capacity}---the available void volume---and not by
kinetics. The outcome then depends only on how the pore geometry evolves with
coating thickness, together with a single number describing how much solute
the exposed caps retain. The purpose of this paper is to turn this intuition
into a quantitative, general theory and to test it.

We define $\Pval(t)$ as the fraction of solute that ends up in the valley
region (the solute column integrated over the footprint of the three-sphere
interstice, normalised by the total solute in one unit cell) and
$\Pcap = 1-\Pval$ as the complementary cap fraction. We show that $\Pval$ can
be built from three ingredients that are treated generally, for any sphere
radius $R$, conformal film thickness $t$, and wetting liquid characterised by
its surface tension $\gamma$, viscosity $\mu$, and equilibrium contact angle
$\theta$: (i)~the exact three-sphere geometry, which fixes a purely geometric,
\emph{size-independent} threshold in the reduced thickness $\tr = t/R$ and
yields the void volumes in closed form; (ii)~Poiseuille flow through the
constricted, nearly triangular throat, which quantifies how fast the liquid
fills the void; and (iii)~thin-film dynamics of the draining solvent film,
which confirms the direction and ordering of drainage. The result is a
piecewise-analytic master equation with a single empirical parameter and a
sharp step that describes the final partitioning of the solute.

Throughout, we keep the theory symbolic and dimensionless so that the results
apply to any hexagonal sphere monolayer. We reduce all lengths by $R$, all
volumes by $R^{3}$, and group the material and process parameters into a single
capillary--drying number. Only in \cref{sec:application} do we insert concrete
numbers, for the experimentally relevant system of polystyrene (PS) beads coated
with tantalum pentoxide (\mbox{Ta$_2$O$_5$}) and wet by ethanol; the
validation checks of \cref{sec:numerical} were carried out for the same system.

The paper is organised as follows. \Cref{sec:geometry} establishes the general
three-sphere geometry, the universal threshold and the exact void volumes.
\Cref{sec:analytics} derives the wicking rate through the throat, the
dimensionless wicking/drying competition, and the capacity-limited master
equation. \Cref{sec:numerical} presents three independent checks of the model:
two numerical (void-volume integration and thin-film simulations) and one
semi-analytic (the wicking/drying balance with the exact throat flow).
\Cref{sec:application} turns the theory to the PS/\mbox{Ta$_2$O$_5$}/ethanol
system and reports all numbers. \Cref{sec:discussion} discusses scope and
implications, including a comparison with measured Purcell factors, and
\cref{sec:conclusion} summarises the main results.

\section{General three-sphere geometry and the universal threshold}
\label{sec:geometry}

\subsection{Setup and reduced variables}
\label{sec:setup}

Consider a two-dimensional hexagonally close-packed monolayer of spheres of
radius $R$ resting on a flat substrate at $z=0$, so that the sphere centres lie
at $z=R$. We take the film to be fully conformal, as produced by ALD: it has the
same thickness $t$ on every exposed surface, so that each sphere is inflated to
an outer radius $R'=R+t$ and the substrate between the spheres is covered by a
layer of thickness $t$. The films of Ref.~\cite{Falamas2026} were grown by
electron-beam evaporation, but the optical response of their upper cups was
found to be best described by such a conformal shell, which is the geometry
adopted here. Any three mutually tangent spheres have centres at the vertices
of an equilateral triangle of side $2R$. We work in the natural reduced
variables
\begin{equation}
	\tr \equiv \frac{t}{R}, \qquad
	\vvoid \equiv \frac{\Vvoid}{R^{3}},
	\label{eq:reduced}
\end{equation}
so that every geometric result below is a pure function of $\tr$, independent of
the absolute sphere size. This is the central simplification exploited
throughout.

The distance $a$ from the centroid of the three-sphere triangle to each vertex
(its circumradius) follows from elementary triangle geometry
(\cref{fig:geometry}a): the centroid divides each median $2\!:\!1$ from the
vertex, and the median of an equilateral triangle of side $s=2R$ has length
$s\sqrt{3}/2$, so $a=\tfrac{2}{3}\cdot s\sqrt{3}/2 = s/\sqrt{3}$. Hence
\begin{equation}
	a = \frac{2R}{\sqrt{3}}, \qquad \frac{a}{R} = \frac{2}{\sqrt{3}}
	\approx 1.1547 .
	\label{eq:a}
\end{equation}
We tile the hexagonal lattice with a rectangular unit cell
$L_x\times L_y = 2R\times 2R\sqrt{3}$ containing two spheres and four
equivalent three-sphere interstices (``centroids''); volumes are reported per
centroid unless stated otherwise.

\subsection{Clearance function and the location of the throat}
\label{sec:throat}

\begin{figure*}[t]
	\includegraphics[width=\linewidth]{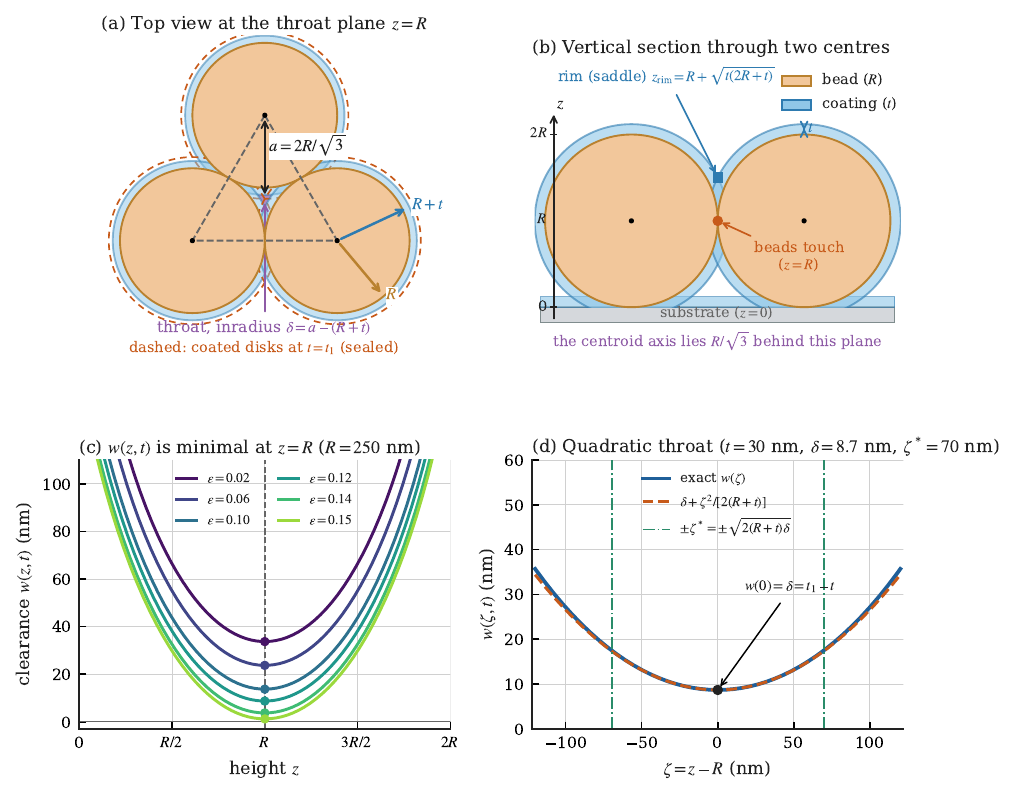}
	\caption{%
		\textbf{Three-sphere geometry (universal in $\tr=t/R$).}
		(a)~Top view down the centroid axis at the throat plane $z=R$: the three
		bare spheres (radius $R$) are mutually tangent; the conformal coating
		(radius $R+t$) leaves a residual curvilinear-triangle throat (shaded)
		of inradius $\delta = a-(R+t)$. As $t\to t_1$ the coated disks meet at
		the centroid (dashed) and the pore seals.
		(b)~Vertical section through the plane of two sphere centres. The film
		is conformal: it covers the whole spheres and the substrate; it
		bridges the two spheres above the equator, forming the rim (saddle) at
		$\zri=R+\sqrt{t(2R+t)}$. The centroid axis (the throat) lies a distance
		$R/\sqrt{3}$ behind this plane, which is why the three-sphere character
		cannot be captured by any single vertical section.
		(c)~Clearance $w(z,t)$ versus height for several $\tr$; the minimum is
		always at $z=R$.
		(d)~Quadratic opening of the throat: exact $w$ versus the parabola of
		\cref{eq:wquad}, with $\pm\zeta^{*}$ of \cref{eq:zeta_star}.
	}
	\label{fig:geometry}
\end{figure*}

At height $z$, a coated sphere centred a horizontal distance $a$ from the
centroid axis presents a circular cross-section of radius
\begin{equation}
	r(z,t) = \sqrt{(R+t)^{2}-(z-R)^{2}} .
	\label{eq:rz}
\end{equation}
The clearance of the void channel, measured from the centroid axis toward one
sphere, is therefore
\begin{equation}
	w(z,t) = a - r(z,t) .
	\label{eq:wz}
\end{equation}
Differentiating, $\partial_z w = (z-R)/r$, which vanishes at $z=R$; the second
derivative there is positive, so \emph{the channel is narrowest at the
equatorial plane} $z=R$, not at the substrate. Indeed,
$w(t,t) = a - \sqrt{4Rt} > a-(R+t) = w(R,t)$ for all $\tr<\trc$. Physically,
the throat sits at the sphere equator and the channel \emph{widens downward}
toward the coated substrate. This single geometric fact---the throat is at the
top of the cup, not at its bottom---is what forbids any re-opening of the void
once the pore is sealed (\cref{sec:regimes}), and hence any non-monotonic
``V-shaped'' recovery of $\Pval$. \Cref{fig:geometry}(c) shows that the minimum
of $w$ stays pinned at $z=R$ for every thickness.

The throat clearance---the inradius of the residual opening at $z=R$, and the
smallest clearance in the whole interstice---is
\begin{equation}
	\delta(t) \equiv w(R,t) = a-(R+t) = t_1 - t ,
	\label{eq:delta}
\end{equation}
which vanishes at the \emph{closing thickness}
\begin{equation}
	\boxed{\; t_1 = R\!\left(\frac{2}{\sqrt{3}}-1\right),
		\qquad
		\trc \equiv \frac{t_1}{R} = \frac{2}{\sqrt{3}}-1 \approx 0.1547 . \;}
	\label{eq:t1}
\end{equation}
The threshold in reduced units, $\trc$, depends only on the hexagonal packing
and not on the sphere size, the film material, or the liquid. For $\tr<\trc$
the throat is open ($\delta>0$); for $\tr>\trc$ the three coated spheres
overlap at the centroid and the under-cup is sealed (\cref{fig:geometry}a).

\subsection{Quadratic opening near the throat}
\label{sec:quadratic}

The shape of the constriction controls the hydraulic resistance of the throat
(\cref{sec:analytics}). For $|\zeta|\ll R+t$ with $\zeta\equiv z-R$, expanding
the square root in \cref{eq:wz} gives a parabola,
\begin{equation}
	w(\zeta,t) \simeq \delta(t) + \frac{\zeta^{2}}{2(R+t)} ,
	\label{eq:wquad}
\end{equation}
whose curvature is set by the coated radius, which tends to $a$ at threshold.
The clearance doubles from $\delta$ to $2\delta$ over the constriction length
\begin{equation}
	\zeta^{*} = \sqrt{2(R+t)\,\delta} .
	\label{eq:zeta_star}
\end{equation}
The parabola is excellent near the throat (\cref{fig:geometry}d) and underlies
the $\delta^{5/2}$ scaling derived in \cref{sec:wicking}.

\subsection{Exact void volumes and the two regimes}
\label{sec:regimes}

\begin{figure*}[t]
	\includegraphics[width=\linewidth]{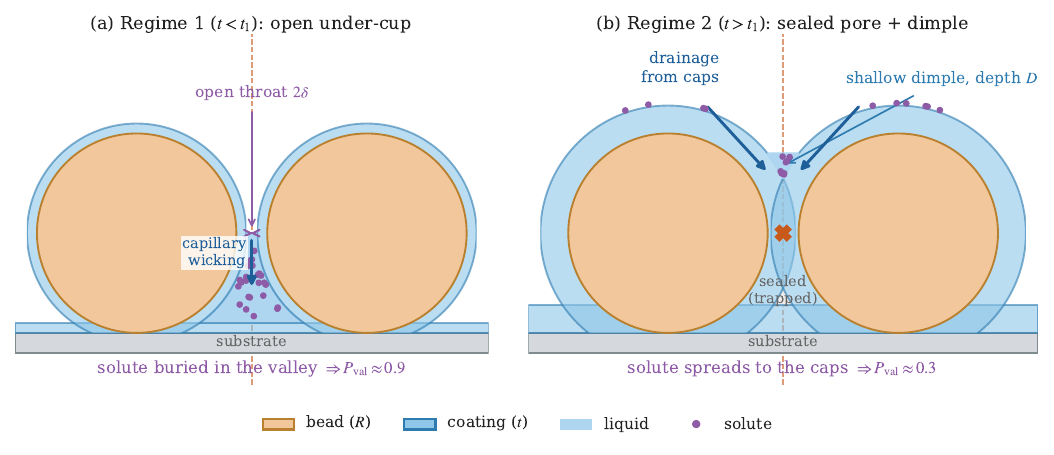}
	\caption{%
		\textbf{The two capillary regimes.}
		Schematic radial section along the centroid axis. The distance from the
		axis to each of the three mutually tangent spheres is $a$; the conformal
		film covers the spheres (radius $R+t$) and the substrate (thickness
		$t$). The throat clearance at $z=R$ is $\delta=a-(R+t)$
		[\cref{eq:wz,eq:delta}].
		(a)~Regime~1 ($\tr<\trc$): the throat is open; the buried under-cup,
		bounded below by the coated substrate, is a deep reservoir filled by
		capillary wicking through the throat, and the solute is buried in the
		valley, giving a large $\Pval$.
		(b)~Regime~2 ($\tr>\trc$): the coated spheres overlap at the centroid,
		sealing the pore; the trapped cavity is inaccessible and only a shallow
		surface dimple (depth $D$) survives, fed by drainage from the cap tops,
		so the solute spreads and $\Pval$ is small.
	}
	\label{fig:regimes}
\end{figure*}

At height $z$ the void of one interstice is the part of the equilateral
triangle of side $2R$ that lies outside the three coated disks of radius $r$
[\cref{eq:rz}] centred on its vertices. Each disk covers a $60^{\circ}$ sector
of the triangle; for $r>R$ neighbouring disks overlap in a lens, half of which
lies inside the triangle; and no point of the triangle is covered by all three
disks as long as $r<a$. Inclusion--exclusion then gives the cross-sectional
void area in closed form,
\begin{equation}
	A(r) = \sqrt{3}\,R^{2} - \frac{\pi}{2}\,r^{2}
	+ 3\,\Theta(r-R)\Bigl[r^{2}\arccos\frac{R}{r} - R\sqrt{r^{2}-R^{2}}\Bigr],
	\qquad r<a ,
	\label{eq:Avoid}
\end{equation}
with $A=0$ for $r\ge a$, and $\Theta$ the Heaviside function. The volumes follow
from a single quadrature in $z$, so that the whole construction is exact; it is
validated against an independent pixel count of the periodic unit cell in
\cref{sec:checkA}. The threshold $\trc$ separates two physically distinct
geometries, sketched in \cref{fig:regimes}.

\paragraph{Regime~1: open under-cup ($\tr<\trc$).}
The interstice is a buried cavity, bounded below by the coated substrate at
$z=t$, connected to the outer surface through the open throat at $z=R$. Its
volume is
\begin{equation}
	\Vcup(t) = \int_{t}^{R} A\bigl(r(z,t)\bigr)\,\mathrm{d}z,
	\qquad
	\vcup(\tr) = \frac{\Vcup}{R^{3}} .
	\label{eq:Vcup}
\end{equation}
$\vcup(\tr)$ decreases monotonically as the spheres inflate and the substrate
film rises, and is a universal function of $\tr$.

\paragraph{Regime~2: sealed cup and surface dimple ($\tr>\trc$).}
Once the throat seals, the buried cavity is closed off and becomes
inaccessible to the liquid. A shallow topographic dimple remains on the outer
surface, bounded below by its floor (the tripoint where the three coated
envelopes meet on the centroid axis) and above by its rim (the two-sphere
saddle). Requiring the centroid axis to touch a coated sphere gives the floor
height, and the two-sphere saddle gives the rim height,
\begin{align}
	\zfl(t) &= R + \sqrt{(R+t)^{2}-a^{2}}, \label{eq:zfloor}\\
	\zri(t) &= R + \sqrt{t(2R+t)} . \label{eq:zrim}
\end{align}
The dimple depth $D(t)=\zri-\zfl$ is maximal exactly at threshold. At
$\tr=\trc$ one has $R+t_1=a$, so $\zfl(t_1)=R$ and, using
$t_1(2R+t_1)=R^{2}(\tfrac{2}{\sqrt{3}}-1)(\tfrac{2}{\sqrt{3}}+1)=R^{2}/3$,
\begin{equation}
	\boxed{\; D(t_1) = \frac{R}{\sqrt{3}} \;}\, ,
	\label{eq:Dmax}
\end{equation}
after which $D(t)$ decreases monotonically to zero (\cref{fig:lub_dimple}b).
The dimple volume
\begin{equation}
	\Vdimple(t)=\int_{\zfl}^{\zri}A\bigl(r(z,t)\bigr)\,\mathrm{d}z ,
	\qquad \vdimple(\tr)=\frac{\Vdimple}{R^{3}},
	\label{eq:Vdimple}
\end{equation}
is likewise monotonically decreasing. At threshold the accessible void
collapses from the cup to the much smaller dimple: the limiting ratio
$\vcup(\trc^{-})/\vdimple(\trc^{+})\approx\numRatioLimit$, taken as
$\tr\to\trc$ from either side, is a universal number (independent of $R$), and
drives the step in $\Pval$ derived in \cref{sec:master}. At any finite
distance from threshold the ratio is larger, since $\vcup$ decreases and
$\vdimple$ increases as $\tr\to\trc$.

\section{Capacity-limited master equation for $\Pval$}
\label{sec:analytics}

The geometry of \cref{sec:geometry} tells us \emph{where} the void volume is,
but not \emph{whether} the liquid has time to reach it before the solvent dries.
We now show that, for realistic liquids and drying times, the answer is
overwhelmingly yes: the filling is \emph{capacity-limited}, and $\Pval$ reduces
to a ratio of void volumes.

\subsection{Wicking through the throat: a nearly triangular duct}
\label{sec:wicking}

Consider Regime~1. The liquid must flow from the outer film, through the throat
at $z=R$, into the under-cup. Near closure the cross-section of the channel at
height $\zeta$ is a small curvilinear triangle bounded by three arcs of radius
$r\simeq a\gg w$, so its sides are nearly straight: it is close to an
equilateral triangle of inradius $w(\zeta)$ and side $s=2\sqrt{3}\,w$. (One can
check that the distance from the axis to the cusps where neighbouring arcs meet
is $2w$ to the same order, as for an equilateral triangle.) The channel is
therefore a slowly varying, nearly triangular duct, and two classical results
apply.

\emph{Driving pressure.} A perfectly wetting liquid invading a duct of
perimeter $\Perim$ and area $A$ gains the surface energy
$\gamma\cos\theta\,\Perim$ per unit length filled, which corresponds to a
capillary suction~\cite{deGennes2004}
\begin{equation}
	\dPcap = \gamma\cos\theta\,\frac{\Perim}{A}
	\simeq \frac{2\gamma\cos\theta}{\delta},
	\label{eq:DeltaP}
\end{equation}
where the last form holds for an equilateral triangle of inradius $\delta$
($\Perim/A=2/\delta$). Equation~\eqref{eq:DeltaP} assumes a wetting liquid,
$\theta<90^\circ$; as $\theta\to 90^\circ$, $\dPcap\to 0$ and wicking ceases.
For ethanol on \mbox{Ta$_2$O$_5$}, $\theta\simeq 0$ and $\cos\theta\approx 1$,
which is the case of \cref{sec:application}.

\begin{figure*}[t]
	\includegraphics[width=\linewidth]{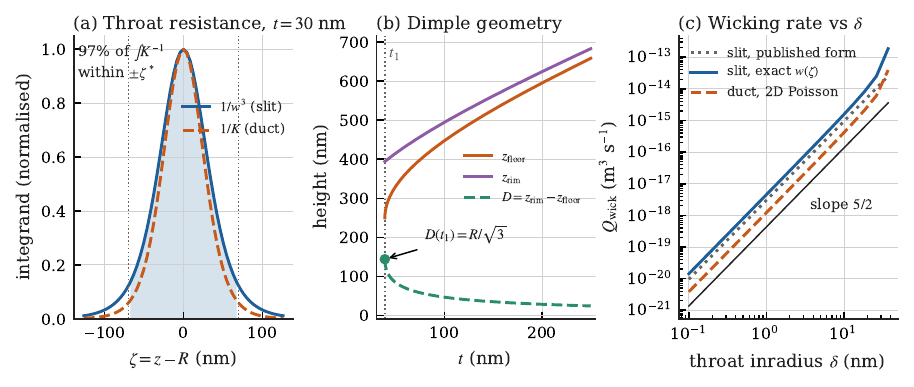}
	\caption{%
		\textbf{Throat resistance, dimple geometry, and wicking scaling.}
		(a)~Integrands of the throat resistance at $t=\SI{30}{nm}$, normalised:
		$1/K(\zeta)$ for the exact cross-sections (duct, dashed) and the slit
		estimate $1/w^{3}$ (solid); both are sharply peaked at the throat, and
		the shaded band $|\zeta|\le\zeta^{*}$ carries most of the integral.
		(b)~Dimple geometry: floor $\zfl(t)$, rim $\zri(t)$, and depth $D(t)$;
		$D$ is maximal ($R/\sqrt{3}$) at $t_1$ and decreases monotonically.
		(c)~Wicking rate versus throat inradius $\delta$ on logarithmic axes for
		the duct model (numerical, exact cross-sections), compared with the
		slit estimates discussed in \cref{sec:discussion}; the thin line has
		slope $5/2$.
	}
	\label{fig:lub_dimple}
\end{figure*}

\emph{Resistance.} For Stokes flow along a slowly varying duct the volumetric
rate is $Q=-K\,\partial_\zeta P/\mu$, where $K$ is the hydraulic conductance of
the local cross-section, obtained from the Poisson problem
$\nabla_\perp^{2}u=-1$ with $u=0$ on the walls, $K=\int u\,\mathrm{d}A$. For an
equilateral triangle of side $s$, $K=\sqrt{3}\,s^{4}/320$~\cite{PatzekSilin2001},
i.e.\ $K=\tfrac{9\sqrt{3}}{20}\,w^{4}$. The resistance of the constriction is
\begin{equation}
	\Rduct = \mu\int\frac{\mathrm{d}\zeta}{K(\zeta)}
	= \frac{20\,\mu}{9\sqrt{3}}\int\frac{\mathrm{d}\zeta}{w(\zeta)^{4}}
	= \frac{20}{9\sqrt{3}}\,\frac{5\pi}{16}\,
	\frac{\mu\sqrt{2(R+t)}}{\delta^{7/2}} ,
	\label{eq:Rduct}
\end{equation}
where the last step uses \cref{eq:wquad} with $u=\zeta/\zeta^{*}$ and
$\int_{-\infty}^{\infty}(1+u^{2})^{-4}\mathrm{d}u=5\pi/16$. The integrand
$w^{-4}$ is sharply peaked at the throat (\cref{fig:lub_dimple}a), so the
integral converges within a few $\zeta^{*}$ and the limits may be taken to
infinity. Hence
\begin{equation}
	\Qwick = \frac{\dPcap}{\Rduct}
	= \frac{36\sqrt{6}}{25\pi}\,
	\frac{\gamma\cos\theta}{\mu\sqrt{R+t}}\,\delta^{5/2}
	\simeq \Cd\,\frac{\gamma\cos\theta}{\mu\sqrt{R}}\,\delta^{5/2},
	\qquad
	\Cd = \frac{36\sqrt{6}}{25\pi}\Bigl(\frac{\sqrt{3}}{2}\Bigr)^{1/2}
	\approx \numCduct ,
	\label{eq:Qwick}
\end{equation}
where the last form uses $R+t\to a$ near threshold. The exponent $5/2$
combines the suction $\propto\delta^{-1}$ and the resistance
$\propto\delta^{-7/2}$. \Cref{eq:Qwick} is asymptotically exact as
$\delta\to 0$: solving the Poisson problem numerically on the exact
cross-sections [\cref{eq:Avoid}] and integrating the resistance over the whole
closed part of the channel gives $\Cd\to\numCzero$ as $\delta\to0$ and
$\Cd=\numCeff$ at $t=\SI{30}{nm}$ (\cref{sec:checkC}, \cref{fig:lub_dimple}c).
The corresponding filling time is
\begin{equation}
	\twick(t) = \frac{\Vcup(t)}{\Qwick(t)}
	\simeq \frac{\mu\sqrt{R}}{\Cd\,\gamma\cos\theta}\,
	\frac{\Vcup(t)}{\delta(t)^{5/2}} .
	\label{eq:twick}
\end{equation}

\subsection{The capillary--drying number and the capacity limit}
\label{sec:capacity}

Whether wicking outruns drying is decided by the ratio $\tdry/\twick$, where
$\tdry$ is the solvent drying time (for spin-coating,
$\tdry\sim\SI{0.1}{s}$~\cite{Meyerhofer1978}; centrifugal forces are negligible
at nanometric gaps~\cite{Emslie1958}). Inserting the reduced variables
$\delta=R(\trc-\tr)$ and $\Vcup=\vcup R^{3}$ into \cref{eq:twick} and forming
the ratio yields a compact, dimensionless result:
\begin{equation}
	\frac{\tdry}{\twick}(\tr)
	= \Cd\;\Cnum\;\frac{(\trc-\tr)^{5/2}}{\vcup(\tr)},
	\qquad
	\boxed{\;\Cnum \equiv \frac{\gamma\cos\theta\,\tdry}{\mu R}\; } .
	\label{eq:ratio}
\end{equation}
All material and process parameters collapse into the single
\emph{capillary--drying number} $\Cnum$, while the $\tr$-dependence is universal.
The larger $\Cnum$, the wider the window of capacity-limited filling.
Equation~\eqref{eq:ratio} shows that wicking dominates ($\tdry/\twick\gg1$)
across essentially the whole open regime and only fails in a narrow band as
$\tr\to\trc$ (where $\delta\to0$); we quantify this window for the
illustrative system in \cref{sec:application}. An analogous, and even more
lopsided, estimate holds in Regime~2. There the driving pressure for drainage
from the cap tops into the dimple is $\dPdrain\sim2\gamma/(R+t)$~\cite{deGennes2004},
and lubricating flow from a cap catchment of area $\sim R^{2}$ with residual film
$h_0$ delivers a volume $\sim\gamma h_0^{3}\,\tdry/(3\mu R)$ per centroid over the
drying time~\cite{Oron1997,Craster2009}, which for any realistic $h_0$ far exceeds
the small dimple volume $\Vdimple$. Both regimes are therefore capacity-limited.

\subsection{Master equation}
\label{sec:master}

When the filling is capacity-limited, the solute distribution at the end of
drying is set purely by the available void volume. $\Pval$ is then the
volumetric fraction of solute-bearing void,
\begin{equation}
	\boxed{\;
		\Pval(\tr) = \frac{\Vvoid(\tr)}{\Vvoid(\tr)+\Vcappar}
		= \frac{\vvoid(\tr)}{\vvoid(\tr)+\vcappar}\; ,
		\label{eq:Pval}
	}
\end{equation}
with
\begin{equation}
	\Vvoid = \begin{cases}
		\Vcup(\tr), & 0<\tr<\trc \quad(\text{Regime 1}),\\[1ex]
		\Vdimple(\tr), & \trc<\tr\lesssim\tr_{\mathrm{flat}}
		\quad(\text{Regime 2}),\\[1ex]
		0, & \tr\gtrsim\tr_{\mathrm{flat}},
	\end{cases}
\end{equation}
where $\vcappar=\Vcappar/R^{3}$ is the reduced solute volume retained on the
caps per centroid---the \emph{sole free parameter}, fixed by the deposition
conditions---and $\tr_{\mathrm{flat}}$ is a soft upper cutoff reached when the
dimple becomes shallower than the residual film and can no longer pin the
receding meniscus. It is a one-parameter fit, not a thickness-dependent
adjustment. Physically $\Vcappar$ is the residual solute left on the exposed
caps after the solvent has drained (a leftover film, contact-line pinning). For
$R=\SI{250}{nm}$, the value $\Vcappar=5\times10^5\nm^3$ used in
\cref{sec:application} ($\vcappar\approx\numVcapRed$) corresponds to a residual
film of only a few nanometres over a cap area of order $\pi R^{2}$. One
measurement of $\Pval$ at a known $t$ fixes $\vcappar$; thereafter
\cref{eq:Pval} predicts the whole curve.

Equation~\eqref{eq:Pval} rests on two implicit assumptions, which we state
explicitly. First, the liquid supply per centroid must be at least
$\Vvoid+\Vcappar$, so that the void is completely filled before the free
surface detaches from the caps. Expressed as an equivalent film thickness over
the centroid footprint $\sqrt{3}R^{2}$, this requires
$h_{\mathrm{liq}}\gtrsim(\Vvoid+\Vcappar)/(\sqrt{3}R^{2})$; for
$R=\SI{250}{nm}$ this is at most $\approx\SI{100}{nm}$, far below the
micrometre-scale wet films produced at the start of spin-coating. If less
liquid were available, the void would be only partly filled and $\Pval$ would
fall below \cref{eq:Pval}. Second, the solute is taken to be uniformly
distributed in the liquid when the drainage pattern freezes, so that it
partitions in proportion to liquid volume; solute accumulation at the receding
contact line (coffee-ring transport) would bias $\Pval$ toward the last-drying
regions and is not included.

\begin{figure*}[t]
	\includegraphics[width=\linewidth]{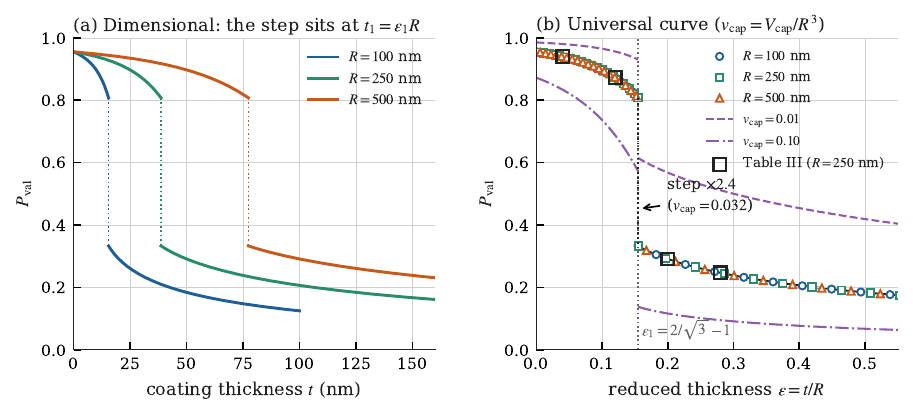}
	\caption{%
		\textbf{Universal master curve for $\Pval$.}
		Because both void volumes scale as $R^{3}$, the reduced volumes
		$\vvoid(\tr)=\Vvoid/R^{3}$ depend on $\tr=t/R$ alone, and
		$\Pval=\vvoid/(\vvoid+\vcappar)$ is universal in $(\tr,\vcappar)$.
		(a)~Plotted against the dimensional thickness $t$ for three sphere
		radii, the step sits at $t_1=\trc R$ and shifts with $R$.
		(b)~The same three data sets coincide on one universal curve versus
		$\tr$ (symbols; line to guide the eye). Open squares are the
		PS/\mbox{Ta$_2$O$_5$} values of \cref{tab:Pval} for $R=\SI{250}{nm}$.
		Purple curves show the dependence on the reduced cap volume
		$\vcappar$. The threshold $\trc=2/\sqrt{3}-1\approx0.155$ depends only
		on the hexagonal packing; the step factor at $\trc$ depends on
		$\vcappar$.
	}
	\label{fig:master}
\end{figure*}

Written in reduced variables, \cref{eq:Pval} is \emph{universal}: for a given
$\vcappar$ the entire $\Pval(\tr)$ curve is independent of the sphere size,
because both void volumes scale as $R^{3}$. \Cref{fig:master} illustrates this:
curves computed for different radii $R$ sit at different \emph{dimensional}
thresholds $t_1=\trc R$ (\cref{fig:master}a) but coincide when plotted against
$\tr$ (\cref{fig:master}b).

Three features of the master curve are worth emphasising.
(i)~In Regime~1, $\Pval$ is large and decreases slowly as $\vcup$ shrinks.
(ii)~At $\trc$ there is a \emph{downward step}, by a factor of
$\numStep$ for $\vcappar=\numVcapRed$: the void-volume collapse
($\vcup(\trc^{-})/\vdimple(\trc^{+})\approx\numRatioLimit$) is compressed by the
$\Vvoid/(\Vvoid+\Vcappar)$ saturation into a smaller change in $\Pval$. The
size of the step depends on $\vcappar$; only the volume ratio is universal.
(iii)~Because $\vdimple(\tr)$ decreases monotonically for $\tr>\trc$, $\Pval$ can
only decrease: \emph{no V-shaped recovery is possible}, a direct consequence of
the throat sitting at the equator (\cref{sec:throat}).

\section{Validation checks}
\label{sec:numerical}

We tested the theory with three independent checks, implemented in Python
scripts. The checks were run for the illustrative
PS/\mbox{Ta$_2$O$_5$}/ethanol system of \cref{sec:application}, but they
validate the general construction: (A)~the void volumes that enter the master
equation, by direct numerical integration; (B)~the direction and $t$-ordering
of drainage, by thin-film simulations using JAX~\cite{jax2018github}; and
(C)~the capacity-limit assumption, by a semi-analytic evaluation of
\cref{eq:ratio} with the exact throat flow and the exact volumes of Check~A.
Checks~A and~B are numerical; Check~C is not an independent simulation but a
quantitative evaluation of the wicking model across the whole open regime.

\subsection{Check A: exact void volumes}
\label{sec:checkA}

\begin{figure}[t]
	\includegraphics[width=\linewidth]{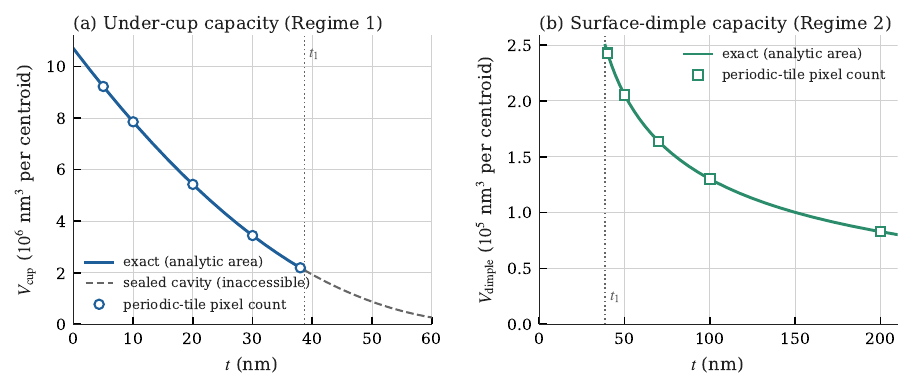}
	\caption{%
		\textbf{Check A: void volumes per centroid ($R=\SI{250}{nm}$).}
		Lines: exact volumes from \cref{eq:Avoid}; symbols: independent pixel
		count of the periodic unit cell.
		(a)~Under-cup volume (Regime~1); dashed: volume of the sealed,
		inaccessible cavity for $t>t_1$.
		(b)~Dimple volume (Regime~2); the steep initial drop marks the rapid
		loss of dimple depth just above threshold.
	}
	\label{fig:checkA}
\end{figure}

The volumes of \cref{eq:Vcup,eq:Vdimple} were obtained by adaptive quadrature of
the closed-form area \cref{eq:Avoid} (relative tolerance $10^{-11}$). As an
independent test that does not use \cref{eq:Avoid} or the reduction to a single
triangle, we counted void pixels in the periodic rectangular unit cell (two
spheres and their periodic images, $2000\times3464$ pixels, 1500 slices in $z$)
and divided by the four centroids per cell. The largest relative difference
between the two methods over the thicknesses of \cref{tab:volumes} is
$\numPixelDev$. \Cref{fig:checkA}
shows the result. $\vcup(\tr)$ decreases monotonically for $\tr<\trc$; above
threshold the cavity below the closure is sealed (dashed), and the accessible
void is the dimple, whose volume decreases monotonically with a steep initial
drop reflecting the large dimple depth just past threshold.

\subsection{Check B: thin-film drainage simulations}
\label{sec:checkB}

Check~B verifies that a draining liquid film actually pools in the
depressions, with the correct ordering in $t$. The lubrication (thin-film)
equation on the coated topography $H(x,y)$ is~\cite{Oron1997,Craster2009}
\begin{equation}
	\partial_t h = \nabla\!\cdot\!\left(\frac{h^{3}}{3\mu}\nabla P\right)
	- J_{\mathrm{evap}},
	\quad
	P = -\gamma\nabla^{2}(h+H) + \Pi(h),
	\label{eq:TFE}
\end{equation}
where $h$ is the film thickness, $\Pi(h)=A_{\mathrm{d}}[(h_*/h)^{3}-1]$ is a
disjoining-pressure regularisation with precursor thickness
$h_*=\SI{2}{nm}$~\cite{Schwartz1998}, and $J_{\mathrm{evap}}$ is a spatially
varying evaporation rate. We take
$J_{\mathrm{evap}}=J_0\bigl[1+\alpha\tilde H(x,y)\bigr]$, where
$\tilde H=(H-H_{\min})/(H_{\max}-H_{\min})$ is the topography min-max
normalised to $[0,1]$, so that evaporation is faster on the caps
($\tilde H=1$) than in the dimples ($\tilde H=0$); we use $\alpha=1$, and the
drainage \emph{direction} (caps to valleys) does not depend on its precise
value. Because $|\nabla H|=O(1)$ on the actual sphere caps, violating the
small-slope assumption, we use a smoothed cosine-hexagonal topography with the
lattice periodicity whose relief is the true cap-top-to-floor relief
$(2R+t)-\zfl$ compressed by the fixed factor $\kappa=\numKappa$. The relief of
this proxy topography ranges from $\numReliefPdeMax$ to
$\SI{\numReliefPdeMin}{nm}$ and its slope stays below $\numSlopePdeMax$; it
reproduces the ordering in $t$ of the real surface, not its dimple depth $D$.
Check~B is therefore a \emph{qualitative} validation of the drainage direction
and ordering, while the quantitative $\Pval$ comes from Checks~A and~C.

Equation~\eqref{eq:TFE} was solved in non-dimensional form (lengths in nm,
$\gamma=\mu=1$) on a $128\times222$ Fourier grid with a linearly stabilised
semi-implicit spectral scheme: the fourth-order term is treated implicitly with
a constant mobility equal to the maximum of $h^{3}/3$ and corrected
explicitly~\cite{Zhu1999}. The initial film has a uniform thickness
$h_0=\SI{30}{nm}$ measured from the local coated surface $H(x,y)$. With the time
step $\Delta\tnd=\numDtPde$, halving the step changes the final film by
$\numDtDiff$ (relative $L^{2}$ norm).

\begin{figure*}[t]
	\includegraphics[width=\linewidth]{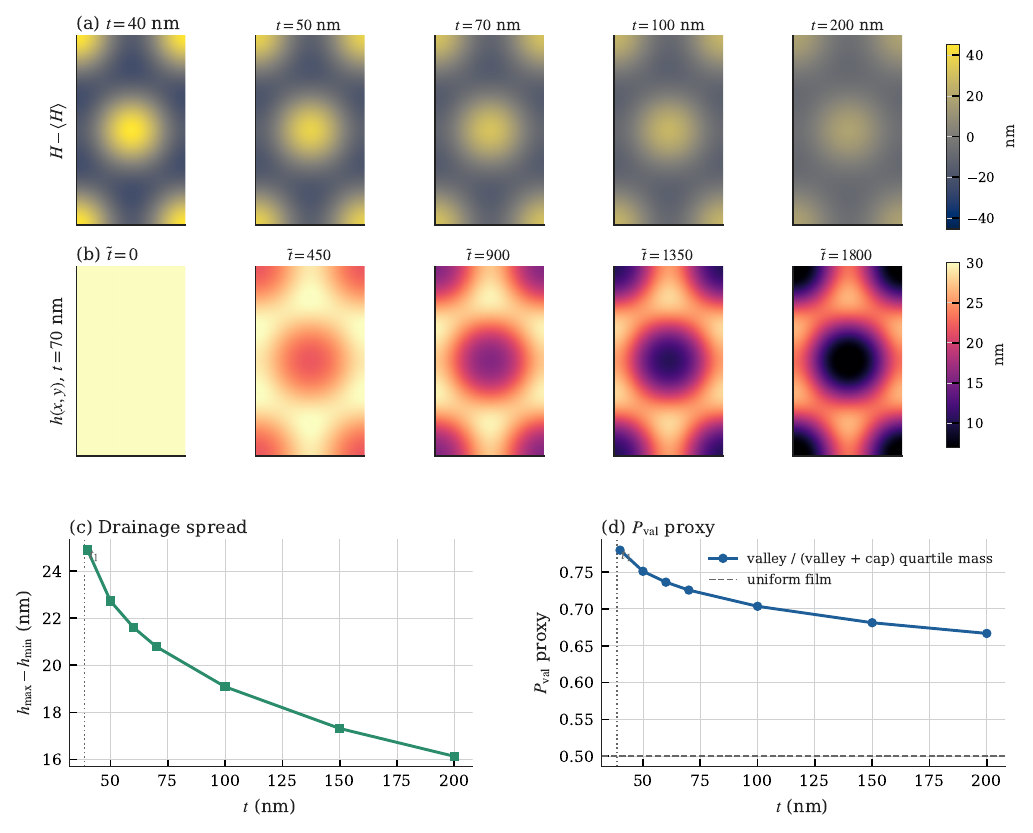}
	\caption{%
		\textbf{Check B, Protocol~1 (PDE integration).}
		(a)~Relief $H-\langle H\rangle$ of the proxy topography for five
		coating thicknesses (common colour scale).
		(b)~Snapshots of $h(x,y,\tnd)$ for $t=\SI{70}{nm}$; the film thins on
		the caps and pools in the depressions.
		(c)~Drainage spread $\hmax-\hmin$ versus $t$.
		(d)~$\Pval$ proxy (valley-quartile mass over valley- plus
		cap-quartile mass) versus $t$; dashed: uniform film.
	}
	\label{fig:checkB_PDE}
\end{figure*}

\emph{Protocol~1 (full PDE).} Integrating \cref{eq:TFE} to non-dimensional time
$\tnd=1800$ (\cref{fig:checkB_PDE}) shows the film thinning on the caps and
pooling in the dimples. The drainage spread $\hmax-\hmin$ decreases
monotonically with $t$, from $\numSpreadPdeMax$ to $\SI{\numSpreadPdeMin}{nm}$
(panel~c), tracking the shrinking topographic relief. A $\Pval$
proxy---the liquid mass over the lowest quartile of $H$ divided by the mass
over the lowest and highest quartiles~\cite{note_proxy}, which equals $0.5$ for
a uniform film---decreases monotonically from $\numProxyMax$ to $\numProxyMin$
(panel~d). The proxy is not comparable with the master-equation $\Pval$: the
smoothed topography has no buried under-cup and its depressions occupy a much
larger area fraction than the true dimples. Explicitly, Check~B establishes only
that the film drains from the caps into the depressions and that this drainage
weakens monotonically with $t$; it validates neither the magnitude of $\Pval$
nor the size of the step at $\trc$, and it does not probe Regime~1.

\begin{figure*}[t]
	\includegraphics[width=\linewidth]{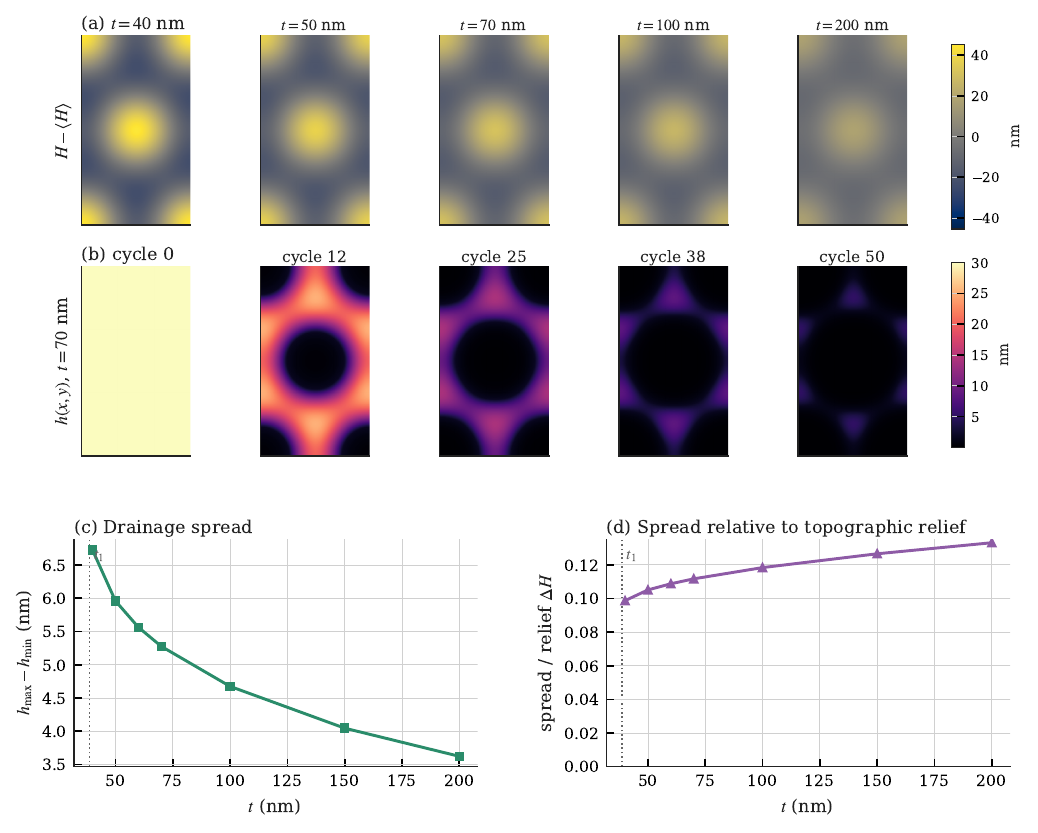}
	\caption{%
		\textbf{Check B, Protocol~2 (capillary-equilibrium cycling).}
		(a)~Relief of the proxy topography for five thicknesses.
		(b)~Snapshots of $h(x,y)$ at cycles $0$--$50$ for $t=\SI{70}{nm}$.
		(c)~Drainage spread versus $t$.
		(d)~Spread divided by the topographic relief $\Delta H$.
	}
	\label{fig:checkB_equil}
\end{figure*}

\emph{Protocol~2 (capillary-equilibrium cycling).} This protocol isolates the
capillary-equilibrium limit of the same problem. Each cycle consists of two
instantaneous operations: (i)~the remaining liquid is redistributed over the
topography so that the free surface is an equipotential,
$h(x,y)+H(x,y)=C$, with $C$ fixed by mass conservation; (ii)~a small fraction of
the film is then removed with the same weight $1+\alpha\tilde H$ as above,
mimicking faster evaporation on the exposed caps. After 50 such cycles
(\cref{fig:checkB_equil}) the drainage spread decreases monotonically with $t$,
from $\numSpreadEqMax$ to $\SI{\numSpreadEqMin}{nm}$, and remains a small
fraction of the topographic relief, rising only slowly from $\numEqRatioMin$ to
$\numEqRatioMax$: at equilibrium the liquid contrast is set by the
topography.

\subsection{Check C: wicking versus drying}
\label{sec:checkC}

\begin{figure*}[t]
	\includegraphics[width=\linewidth]{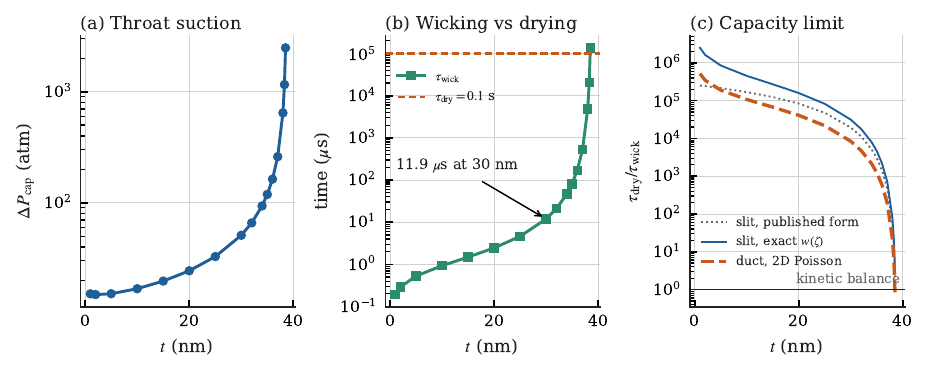}
	\caption{%
		\textbf{Check C: capillary wicking dominates evaporation.}
		(a)~Capillary suction $\dPcap$ at the throat in Regime~1.
		(b)~Wicking time $\twick$ (duct model) versus the drying time $\tdry$.
		(c)~Ratio $\tdry/\twick$ on a logarithmic scale for the duct model
		(thick dashed) and for the two slit estimates of
		\cref{sec:discussion}; the horizontal line marks kinetic balance.
	}
	\label{fig:checkC}
\end{figure*}

We evaluated $\tdry/\twick(\tr)$ across the open regime with the exact throat
flow. At each height the conductance $K(\zeta)$ of the exact curvilinear
cross-section was obtained by solving the Poisson problem of
\cref{sec:wicking} with a second-order finite-difference scheme that treats the
curved walls with unequal arms~\cite{ShortleyWeller1938}; on an equilateral
triangle the scheme reproduces the exact conductance to $\numPoissonErr$ at the
resolution used. The
resistance was integrated over the whole closed part of the channel,
$|\zeta|<\sqrt{t(2R+t)}$, and the suction was taken from the exact perimeter
and area of the throat section [\cref{eq:DeltaP}]. \Cref{fig:checkC} shows the
result. The ratio $\tdry/\twick$ decreases from $\numRatioMax$ at
$t=\SI{1}{nm}$ to $\numRatioChain$ at $t=\SI{30}{nm}$ and crosses kinetic
balance ($\tdry/\twick=1$) only within $\delta\approx\SI{\numDeltaCrit}{nm}$
of threshold. The capacity-limited assumption is thus justified everywhere it
matters, independently of the thin-film approximation of Check~B.

\section{Illustrative application: PS beads, Ta$_2$O$_5$, ethanol}
\label{sec:application}

We now insert concrete numbers for the experimentally relevant system: PS beads
of radius $R=\SI{250}{nm}$, a conformal \mbox{Ta$_2$O$_5$} film of thickness
$t$, and ethanol as the solvent ($\gamma\approx\SI{22}{mN/m}$,
$\cos\theta\approx1$ on \mbox{Ta$_2$O$_5$}~\cite{vanOss1988},
$\mu\approx\numMu$\,Pa\,s at room temperature). The general results of
\cref{sec:geometry,sec:analytics} then become the following.

\paragraph{Threshold.}
From \cref{eq:t1}, $t_1=\trc R\approx\SI{\numTone}{nm}$, and the maximal dimple
depth is $D(t_1)=R/\sqrt{3}\approx\SI{\numDtone}{nm}$.

\paragraph{Void volumes.}
\Cref{tab:volumes} lists the exact volumes (Check~A). At the two tabulated
thicknesses that bracket the threshold,
$\Vcup(\SI{38}{nm})/\Vdimple(\SI{40}{nm})\approx\numRatioTable$; this is
slightly above the limiting ratio $\numRatioLimit$ because neither thickness
lies exactly at $t_1$.

\begin{table}[t]
	\caption{%
		Accessible void volume per centroid (Check~A) for $R=\SI{250}{nm}$:
		under-cup volume in Regime~1, dimple volume in Regime~2, and dimple
		depth $D$.}
	\label{tab:volumes}
	\begin{ruledtabular}
		\begin{tabular}{rcccc}
			$t$ (nm) & $\tr=t/R$ & Regime & $\Vvoid$ (nm$^3$) & $D$ (nm) \\
			\hline
			5 & 0.020 & 1 & $9.23\times10^{6}$ & --- \\
10 & 0.040 & 1 & $7.86\times10^{6}$ & --- \\
20 & 0.080 & 1 & $5.43\times10^{6}$ & --- \\
30 & 0.120 & 1 & $3.44\times10^{6}$ & --- \\
38 & 0.152 & 1 & $2.19\times10^{6}$ & --- \\
40 & 0.160 & 2 & $2.43\times10^{5}$ & 119.3 \\
50 & 0.200 & 2 & $2.05\times10^{5}$ & 84.2 \\
70 & 0.280 & 2 & $1.64\times10^{5}$ & 61.7 \\
100 & 0.400 & 2 & $1.30\times10^{5}$ & 47.0 \\
200 & 0.800 & 2 & $8.29\times10^{4}$ & 29.0

		\end{tabular}
	\end{ruledtabular}
\end{table}

\paragraph{Capillary--drying number and the capacity limit.}
With the above liquid and $\tdry=\SI{0.1}{s}$, the capillary--drying number is
\begin{equation}
	\Cnum = \frac{\gamma\cos\theta\,\tdry}{\mu R}
	\approx \numLambda ,
	\label{eq:Lambda_num}
\end{equation}
an enormous number. \Cref{tab:chain} evaluates the full wicking chain at a
representative $t=\SI{30}{nm}$ ($\delta=\SI{8.7}{nm}$): the wicking time is
$\twick\approx\SI{\numTwick}{\micro s}$, almost four orders of magnitude below
$\tdry$. Across the open regime $\tdry/\twick$ exceeds unity everywhere except
within $\delta\approx\SI{\numDeltaCrit}{nm}$ of threshold
(\cref{fig:checkC}c)---a distance of atomic size, and hence physically
irrelevant.
The illustrative system is therefore firmly capacity-limited, in both regimes.

\begin{table}[t]
	\caption{%
		Wicking chain at $t=\SI{30}{nm}$ for $R=\SI{250}{nm}$, duct model with
		exact cross-sections [\cref{eq:DeltaP,eq:Rduct,eq:Qwick,eq:twick}].}
	\label{tab:chain}
	\begin{ruledtabular}
		\begin{tabular}{lcc}
			Quantity & Expression & Value \\
			\hline
			Throat inradius $\delta$ & $t_1-t$ & 8.7\,nm \\
Constriction length $\zeta^{*}$ & $\sqrt{2(R+t)\delta}$ & 70\,nm \\
Capillary pressure $\Delta P_{\rm cap}$ & $\gamma\cos\theta\,\mathcal{P}/A$ & 51\,atm \\
Resistance $\mathcal{R}$ & $\mu\int\mathrm{d}\zeta/K(\zeta)$ & $1.8\times10^{22}$\,kg\,m$^{-4}$\,s$^{-1}$ \\
Throat perimeter $\mathcal{P}$ & exact (3 arcs) & 95\,nm \\
Throat area $A$ & exact & 404\,nm$^2$ \\
Wicking rate $Q_{\rm wick}$ & $\Delta P_{\rm cap}/\mathcal{R}$ & $2.9\times10^{-16}$\,m$^3$\,s$^{-1}$ \\
Wicking time $\tau_{\rm wick}$ & $V_{\rm cup}/Q_{\rm wick}$ & 11.9\,$\mu$s \\
Drying time $\tau_{\rm dry}$ & (expt.) & 0.1\,s \\
\hline
$\tau_{\rm dry}/\tau_{\rm wick}$ & & $\boldsymbol{\sim8.4\times10^{3}}$

		\end{tabular}
	\end{ruledtabular}
\end{table}

\paragraph{Predicted $\Pval$.}
Using the single fitted parameter $\Vcappar=5\times10^5\nm^3$
($\vcappar\approx\numVcapRed$), the master equation \eqref{eq:Pval} gives the
values in \cref{tab:Pval}: $\Pval\approx{}$\numPvalLo{} in Regime~1, stepping
down to ${\approx}\,$\numPvalHi{} in Regime~2. These are the points overlaid on the
universal curve of \cref{fig:master}(b).

\begin{table}[t]
	\caption{%
		Predicted $\Pval$ for $R=\SI{250}{nm}$, $\Vcappar=5\times10^5\nm^3$.}
	\label{tab:Pval}
	\begin{ruledtabular}
		\begin{tabular}{rccc}
			$t$ (nm) & Regime & $\Vvoid$ (nm$^3$) & $\Pval$ \\
			\hline
			10 & 1 (cup) & $7.86\times10^{6}$ & 0.940 \\
30 & 1 (cup) & $3.44\times10^{6}$ & 0.873 \\
50 & 2 (dimple) & $2.05\times10^{5}$ & 0.291 \\
70 & 2 (dimple) & $1.64\times10^{5}$ & 0.247

		\end{tabular}
	\end{ruledtabular}
\end{table}

\section{Discussion}
\label{sec:discussion}

\paragraph{Throat flow: duct versus slit.}
A simpler estimate treats the throat as a slit of half-width $w(\zeta)$, with a
flow rate per unit width $q=-(2w^{3}/3\mu)\partial_\zeta P$ multiplied by a
wetted perimeter and driven by $\gamma\cos\theta/\delta$. This picture gives the
same exponent $5/2$ but an arbitrary prefactor, because the throat has no
parallel walls: it is a closed, nearly equilateral section whose perimeter at
$t=\SI{30}{nm}$ is $\SI{\numPerimThirty}{nm}$, about $11\delta$. With a
perimeter of $6\delta$ and the integral truncated at $\pm\zeta^{*}$, the slit
estimate gives $\twick\approx\SI{\numTwickPaper}{\micro s}$ at
$t=\SI{30}{nm}$; with the exact clearance and perimeter it gives
$\SI{\numTwickSlit}{\micro s}$, against $\SI{\numTwick}{\micro s}$ for the duct.
All three are shorter than $\tdry$ by about four orders of magnitude
(\cref{fig:checkC}c), so
the capacity limit does not depend on the flow model; the duct model is the
one free of adjustable geometric factors, and it is also the most
conservative, giving the slowest filling.

\paragraph{Scope of the two Check~B protocols.}
The PDE and equilibrium-cycling protocols probe the two complementary limits of
the same draining-film problem. Both give a drainage spread that decreases
monotonically with $t$, and the PDE gives a $\Pval$ proxy above the
uniform-film value of $0.5$ at every thickness. Equation~\eqref{eq:TFE} is
strictly valid only for $|\nabla H|\ll1$, which fails on the actual sphere
caps; both protocols therefore use the smoothed proxy topography and are
qualitative. The \emph{quantitative} prediction rests on the exact void volumes
(Check~A) and the throat flow (Check~C), neither of which invokes the thin-film
approximation.

\paragraph{Solute rheology and contact-angle dynamics.}
In practice the deposition solution may contain a polymer (e.g.\ poly(vinyl
pyrrolidone)) whose concentration, and hence effective viscosity, rises sharply
during drying. Using the late-stage viscosity in \cref{eq:twick} lowers $\Cnum$
and widens the kinetics-limited band from atomic size to a few nanometres near
threshold, softening the sharp step into a sigmoid. It does not move the
threshold, change the monotonicity of $\Pval$, or alter the main conclusions.
The Cox--Voinov~\cite{Cox1986} correction to the contact angle is negligible at
the small capillary numbers of these geometries.

\paragraph{Mode of deposition.}
The theory assumes a conformal film of uniform thickness on the spheres and on
the substrate. The film on the substrate matters in Regime~1: it raises the
bottom of the cup to $z=t$ and removes a layer of the void, which shrinks
$\Vcup$ appreciably near threshold, whereas it leaves the threshold, the
dimple, and Regime~2 unchanged. Strongly directional deposition, whose
thickness falls to zero at the sphere equator, would close the pore at a
different height and thickness and requires a separate geometric analysis;
the conformal description is supported by the optical response of the
evaporated films of Ref.~\cite{Falamas2026}.

\paragraph{Practical implications.}
The theory offers a geometric design rule for placing a dilute solute---a
fluorescent dye, a nanoparticle---either in the interstitial valleys or on the
exposed caps, which is the spatial degree of freedom that controls fluorescence
yield and the ensemble Purcell factor on these lattices~\cite{Falamas2026}.
Because the step sits at the universal $\trc$, the switch can be placed at any
desired absolute thickness simply by choosing the bead radius: $t_1=\trc R$.
Operating below $t_1$ gives strong valley localisation ($\Pval\approx0.9$);
operating above it spreads the solute to the caps ($\Pval\lesssim1/3$). This provides
two distinct surface configurations from the same materials by tuning a single
deposition parameter. Moreover, the sole free parameter $\vcappar$ is fixed by
one measurement of $\Pval$ at any known $t$; thereafter the whole curve is
predicted without further fitting.

\paragraph{Connection to measured Purcell factors.}
The same materials and deposition sequence---PS spheres, a \mbox{Ta$_2$O$_5$}
shell, and a dye spin-coated from ethanol---were used in a recent fluorescence
study of hexagonal microsphere lattices~\cite{Falamas2026}. There, Rhodamine~6G
in poly(vinyl pyrrolidone)/ethanol was deposited on lattices with
$R=\SI{230}{nm}$ and shell thicknesses $t=\SI{10}{nm}$, $30$, $50$ and
$\SI{70}{nm}$. Ensemble Purcell factors extracted from time-resolved decays were
compared with a spatial average over emitters on the caps versus in the valleys
(their Fig.~5 and Sec.~3.5). The on-cap probability was modelled as a logistic
function of $t/R$ and, after a global fit, increased from $\sim 0.05$ at
$\SI{10}{nm}$ to $\sim 0.74$ at $\SI{70}{nm}$.

The present theory supplies first-principles weights. Because $\trc$ is
universal, the closing thickness for those beads is
$t_1=\trc R\approx\SI{\numToneExp}{nm}$: the $\SI{10}{nm}$ and
$\SI{30}{nm}$ shells lie in Regime~1, the $\SI{50}{nm}$ and $\SI{70}{nm}$
shells in Regime~2. With the same reduced cap volume
$\vcappar\approx\numVcapRed$ as in \cref{tab:Pval}, not refitted, the master
equation gives $\Pcap=1-\Pval\approx\numPcapTen$, $\numPcapThirty$,
$\numPcapFifty$ and $\numPcapSeventy$ at $t=\SI{10}{nm}$, $30$, $50$ and
$\SI{70}{nm}$. These values agree with the fitted on-cap probabilities of
Ref.~\cite{Falamas2026} at the two ends of the measured range ($\numPcapTen$
versus $\sim0.05$ at $\SI{10}{nm}$; $\numPcapSeventy$ versus $\sim0.74$ at
$\SI{70}{nm}$). At $30$ and $\SI{50}{nm}$ the reference values are read from the
fitted logistic rather than extracted independently, so the comparison there is
with an interpolant, not with data. Within this limit, the theory replaces the
phenomenological logistic by a geometric step at $\trc$. The poly(vinyl
pyrrolidone) in the experimental formulation raises the late-stage viscosity
and, as noted above, rounds that step into a sigmoid---precisely the shape that
a logistic interpolant would capture. The fluorescence enhancement reported
there peaks at intermediate shells ($\SI{30}{nm}$--$\SI{50}{nm}$), i.e.\ across
the same geometric threshold, where emitters leave the interstitial valleys for
the caps while the lattice resonance still overlaps the dye band.

\section{Conclusion}
\label{sec:conclusion}

We have derived and validated a general, first-principles, piecewise-analytic
expression for $\Pval$, the probability that a solute deposited from a volatile
solvent localises in the interstitial valleys of a conformally coated hexagonal
sphere monolayer. The theory holds for arbitrary sphere radius $R$, conformal
film thickness $t$, and wetting liquid, and reduces to a universal form in the
variables $(\tr,\vcappar)$. The main results are:
\begin{enumerate}

	\item \textbf{Universal geometric threshold.} A size-independent closing
	threshold $\trc=2/\sqrt{3}-1\approx0.155$ separates an open-cup regime from a
	sealed-cup/surface-dimple regime. The throat sits at the sphere equator, so
	the pore closes from the top and the accessible void volume decreases
	monotonically beyond threshold. The dimensional threshold is $t_1=\trc R$.

	\item \textbf{Exact void volumes.} The cross-sectional void area has a
	closed form, so the cup and dimple volumes follow from a single quadrature;
	an independent pixel count of the periodic unit cell agrees to within
	$\numPixelDev$.

	\item \textbf{Capacity-limited filling.} The throat is a nearly triangular
	duct, through which the wicking rate scales as $\delta^{5/2}$ with a
	closed-form prefactor confirmed by exact flow calculations. The competition
	with drying collapses onto a single capillary--drying number
	$\Cnum=\gamma\cos\theta\,\tdry/(\mu R)$; when $\Cnum$ is large the filling is
	capacity-limited and $\Pval$ is a ratio of void volumes, with kinetic
	corrections confined to an inaccessibly narrow band near threshold.

	\item \textbf{Universal master curve and step.} $\Pval$ is large
	($\approx0.9$) in Regime~1, drops at threshold by a factor that depends on
	$\vcappar$ ($\numStep$ for $\vcappar=\numVcapRed$; void-volume ratio
	$\approx\numRatioLimit$), and decreases monotonically thereafter, with no
	V-shaped recovery. In reduced variables the whole curve is the same for all
	sphere sizes.

	\item \textbf{Validation.} Exact void-volume integration with an
	independent pixel count (Check~A), a semi-analytic evaluation of the
	wicking/drying balance with the exact throat flow (Check~C), and
	qualitative thin-film drainage simulations (Check~B) support the geometry,
	the capacity limit, and the drainage direction and ordering, respectively.

	\item \textbf{Single-parameter predictability.} The one free parameter
	$\vcappar$ is set by a single measurement; thereafter the entire
	$\Pval(\tr)$ curve follows without further fitting.

\end{enumerate}

For the experimentally relevant PS/\mbox{Ta$_2$O$_5$}/ethanol system with
$R=\SI{250}{nm}$, the threshold is $t_1\approx\SI{\numTone}{nm}$, the
capillary--drying number is $\Cnum\approx\numLambda$, and $\Pval$ steps from
$\approx0.9$ to $\approx0.3$. For the $R=\SI{230}{nm}$ lattices of
Ref.~\cite{Falamas2026}, the same reduced cap volume places the threshold at
$t_1\approx\SI{\numToneExp}{nm}$ and gives on-cap fractions rising from
$\numPcapTen$ to $\numPcapSeventy$ between $10$ and $\SI{70}{nm}$, in agreement
with the values inferred from measured Purcell factors at both ends of that
range.

\begin{acknowledgments}
	The authors thank the CRPP computing facility for access to numerical
	resources.
\end{acknowledgments}



\begin{thebibliography}{99}
\bibitem{note_proxy}
	The $\Pval$ proxy is $M_{\mathrm{low}}/(M_{\mathrm{low}}+M_{\mathrm{high}})$,
	where $M_{\mathrm{low}}$ ($M_{\mathrm{high}}$) is the liquid mass over the
	quarter of the unit cell with the lowest (highest) topography $H$. It equals
	$0.5$ for a uniform film and increases as liquid drains from the caps into
	the depressions.
\end{thebibliography}
\end{document}